\documentclass[a4paper,11pt]{article}

\usepackage{jheppub} 

\usepackage[T1]{fontenc} 
\usepackage{orcidlink}
\newcommand{\rh}{r_{\text{h}}}
\newcommand{\dd}{\text{d}}

\title{\boldmath Overcharging a nonsingular black hole in general relativity: the nonlinear electrodynamic field effects}

\author[a,b,c]{Wei-Jie Miao\orcidlink{0009-0000-2365-4507}}
\author[a,b,1]{and Si-Jiang Yang\orcidlink{0000-0002-8179-9365}\note{Corresponding author.}}

\affiliation[a]{Lanzhou Center for Theoretical Physics, Key Laboratory of Theoretical Physics of Gansu Province, Key Laboratory of Quantum Theory and Applications of MoE, Gansu Provincial Research Center for Basic Disciplines of Quantum Physics, Lanzhou University, Lanzhou 730000, China}
\affiliation[b]{Institute of Theoretical Physics $\&$ Research Center of Gravitation, School of Physical Science and Technology, Lanzhou University, Lanzhou 730000, China}
\affiliation[c]{Cuiying Honors College, Lanzhou University, Lanzhou 730000, China}

\emailAdd{miaowj21@lzu.edu.cn}
\emailAdd{yangsj@lzu.edu.cn}

\abstract{The Ay\'on-Beato Garc\'ia (ABG) solution describes a nonlinear electrodynamic nonsingular black hole in general relativity and can be regarded as a strong field correction to the Reissner-Nordstr\"om solution. We exam the possibility of destroying the ABG nonsingular black hole with a test charged particle and a complex scalar field. By comparing with the results of gadenken experiment to destroy the Reissner-Nordstr\"om black hole, we get the nonlinear electrodynamic field effects on the destruction of the event horizon. We obtain the parameter regions of the particle and scalar field, which allow us to destroy the event horizon of an extremal and near-extremal ABG black hole. Our findings show that both can be destroyed due to nonlinear electrodynamic effects. Interestingly, the parameter intervals for the charged particle and scalar field needed to destroy the event horizon of an extremal black hole are identical. Since the ABG black hole is nonsingular, our results remain consistent with the weak cosmic censorship conjecture and may offer a means to explore the interior of a black hole.}

\begin{document}
\maketitle
\flushbottom

\section{Introduction}\label{sec:intro}

In Einstein's general relativity, Penrose proved that gravitational collapse of matter satisfying reasonable energy conditions inevitably leads to spacetime singularities. This is the so-called Hawking–Penrose singularity theorem \cite{Penrose:1964wq,Hawking:1970zqf}. To avoid the appearance of ``naked singularities'' and rescue general relativity from the loss of predictability, Penrose proposed the weak cosmic censorship conjecture~\cite{Penrose:1969pc}, which states that singularities generated from gravitational collapse with physically reasonable matter must be covered by an event horizon and can never be seen by distant observers. A general proof of the conjecture is known to be quite challenging and remains beyond reach.

In the pioneer work of testing the weak cosmic censorship conjecture, Wald proposed a gedanken experiment by dropping test particles with large charge or angular momentum into an extremal Kerr-Newman black hole, and found that particles causing the destruction of the event horizon cannot be captured by the extremal black hole due to electromagnetic and centrifugal repulsion force~\cite{Wald:1974hkz}. This work is a kind of evidence for the validity of the weak cosmic censorship conjecture. Systematic studies by Rocha and Cardoso et
al. for Ba\~nados–Teitelboim–Zanelli (BTZ) black holes~\cite{Rocha:2011wp}, rotating quantum BTZ black hole~\cite{Frassino:2024fin},
higher-dimensional Myers–Perry family of rotating black
holes and a large class of five-dimensional black rings~\cite{Bouhmadi-Lopez:2010yjy}
also suggested that extremal black holes cannot be destroyed
by test particles. Nevertheless, further investigations by Hod and Hubeny showed that near-extremal black holes might be destroyed by test particles~\cite{Hubeny:1998ga,Hod:2002pm}. The work of Jacobson and Sotiriou also indicates that a near-extremal Kerr black hole can be destroyed by test particles~\cite{Jacobson:2009kt,Jacobson:2010iu}. Similar counterexamples are found in black holes with NUT parameters or magnetized black holes in general relativity~\cite{Yang:2020iat,Yang:2023hll,Feng:2020tyc,GaGa22} and other black holes in modified theories of gravity~\cite{Liang:2018wzd,Ghosh:2019dzq,Ghosh:2021cub}. In addition, Gao and Zhang's investigation suggests that if the second-order of energy, charge, and angular momentum of particles are taken into account, even an extremal Kerr-Newman black hole can be destroyed~\cite{Gao:2012ca}. Similar results can be found for the $4D$ Einstein-Gauss-Bonnet black hole~\cite{Yang:2020czk}.
Besides destroying the event horizon with test particles, we can also do the gedanken experiment through the scattering of test fields. By shooting a charged scalar field into an extremal magnetized Kerr-Newman black hole,
Semiz found that an extremal dyonic Kerr-Newman black hole cannot be destroyed~\cite{Semiz:2005gs}. Systematic investigations for scalar field scattering by some other black holes suggest the same results~\cite{Gwak:2022mze,Gwak:2021tcl,Gwak:2019rcz,Liang:2020hjz,Zhao:2023vxq,Wang:2019jzz,Bai:2020ieh}. Usually, classical fields cannot destroy the event horizon.
Recently, Sorce and Wald proposed a new thought experiment that has taken into account second-order perturbations from matter fields. The investigation shows that a near-extremal Kerr–Newman black hole cannot be destroyed and the weak cosmic censorship conjecture is valid~\cite{Sorce:2017dst}. The systematic work for other black holes using this new formalism suggests that both extremal and near-extremal black holes cannot be destroyed~\cite{Chen:2019nhv,Lin:2022ndf,Wang:2021kcq,Jiang:2022zod,Zhang:2020txy,Jiang:2020xow,Qu:2021hxh,Qu:2020nac,Shaymatov:2019pmn,Shaymatov:2019del,Shaymatov:2020byu,Wu:2024ucf,Li:2020smq}. Using the Iyer-Wald formalism even to high-order perturbations, the results indicate that the weak cosmic censorship conjecture cannot be violated~\cite{Sang:2021xqj,Wang:2022umx,Wang:2020vpn}.

Spacetime singularities are widely regarded as a sign of the failure of general relativity. When quantum effects near singularities are taken into account, spacetime singularities might disappear. However, a consistent quantum theory of gravity is still elusive. To solve the problem of black hole singularity, many viewpoints have been proposed to obtain nonsingular black holes, such as employing ``black-bounce'' mechanism~\cite{Simpson:2018tsi}, properly modifying the metric~\cite{Capozziello:2024ucm,Ling:2021olm,Feng:2023pfq,Hu:2023iuw} or introducing a nonlinear electrodynamic field~\cite{Dymnikova:2015hka,Fan:2016hvf}. For reviews of nonsingular black holes, refer to Refs.~\cite{Lan:2023cvz,Bambi:2023try,Li:2024rbw}. Recently, the investigation of nonsingular black holes has attracted a lot of attention. By constructing an action for a nonlinear electrodynamic field coupled with gravity, Ay\'on--Beato and García obtained an analytical nonsingular black hole solution in Einstein's general relativity~\cite{ABGa98}. Then Cai and Miao achieved a kind of generalized ABG related black hole solutions
which depend on five parameters named the mass, charge, and three parameters related to the nonlinear electrodynamic field~\cite{Cai:2021ele}.

Since there is no spacetime singularity for regular black holes, the destruction of the event horizon of a nonsingular black hole does not violate the weak cosmic censorship conjecture, and nonsingular black holes are not protected by the weak cosmic censorship conjecture~\cite{Li:2013sea}.
The destruction of the event horizon of nonsingular black holes is an active area of research~\cite{Yang:2022yvq,Jiang:2020mws}. Li and Bambi discussed the feasibility of destroying the horizon of rotating nonsingular Bardeen and Hayward black holes~\cite{Li:2013sea}. The source of the Bardeen black hole has been interpreted as a magnetic monopole, which can be derived from nonlinear electrodynamics~\cite{Ayon-Beato:2000mjt}. It was found that these nonsingular black holes might be destroyed by test particles~\cite{Li:2013sea}. Further work of Liu and Gao strongly supports that nonlinear electrodynamic field corrected black holes can be destroyed by test magnetic or electric charged particles~\cite{LiGa20}.

Motivated by the work of Liu and Gao, we investigate the possibility of destroying the event horizon of the ABG nonsingular black hole with a test particle and a test complex scalar field and explore the nonlinear electrodynamic field effects on the destruction of the event horizon. We get the parameter region for the energy and charge of a test particle and a complex scalar field that can destroy the event horizon. The results show that both an extremal and a near-extremal ABG black hole can be destroyed by a charged test particle and a complex scalar field. Interestingly, we find that the parameter interval to destroy an extremal ABG black hole by a complex scalar field is exactly the same as that of a particle.

The outline of the paper is as follows. In Sec.~\ref{lagkh}, we briefly review the characteristics of the ABG black hole. In Sec.~\ref{particle situation}, we investigate the possibility of overcharging the ABG black hole with a test charged particle. In Sec.~\ref{scalar situation}, we try to destroy the event horizon of an extremal ABG black hole with a charged scalar field. The last section
is devoted to discussion and conclusion.

\section{The nonsingular black hole in general relativity}
\label{lagkh}

Spacetime singularities are windows into physics beyond general relativity. When quantum effects are taken into account, spacetime singularities might disappear. There are many viewpoints to avoid spacetime singularity. One of the ways is to assume gravity coupled to a nonlinear electrodynamic field. In this section, we give a brief introduction to a nonlinear electrodynamic field nonsingular black hole in general relativity, as proposed by Ay\'on--Beato and Garc\'ia.

The action for the gravity theory is given by~\cite{Salazar:1987ap}
\begin{equation}
	\mathcal{S}=\int d^4x\left(\frac{1}{16\pi}R-\frac{1}{4\pi}\mathcal{L}(F)\right),
\end{equation}
where $R$ is the curvature scalar, and $\mathcal{L}$ is the Lagrangian for the nonlinear electrodynamic field, which is a function of the electromagnetic field $F=1/4~F_{\mu\nu}F^{\mu\nu}$. Alternatively, the action for the nonlinear electrodynamic field can be described by the Hamiltonian~\cite{Salazar:1987ap}:
\begin{equation}
	\mathcal{H}=2F\mathcal{L}_F-\mathcal{L},
\end{equation}
where $\mathcal{L_{F}}$ is the derivative of the Lagrangian with respect to $F$.
Defining $P_{\mu\nu}=\mathcal{L}_FF_{\mu\nu}$, then
the Hamiltonian $\mathcal{H}$ is a function of
\begin{equation}
   P=\frac{1}{4}P_{\mu\nu}P^{\mu\nu}=(\mathcal{L}_F)^2F ,
\end{equation}
 which is determined by the nonlinear electrodynamic source:
\begin{equation}
	\mathcal{H}(P)=P\frac{(1-3\sqrt{-2q^2P})}{\left( 1+\sqrt{-2q^2P} \right) ^3}-\frac{3}{2q^2s}\left( \frac{\sqrt{-2q^2P}}{1+\sqrt{-2q^2P}} \right) ^{\frac{5}{2}},	
\end{equation}
where $s=|q|/2m$ and the invariant $P$ are non-negative quantities. The corresponding Lagrangian occurs to be
\begin{equation}
\begin{split}
   \mathcal{L}\left( P \right)  =P\frac{\left( 1-8\sqrt{-2q^2P}-6q^2P \right)}{\left( 1+\sqrt{-2q^2P} \right) ^4} -\frac{3}{4q^2s}\frac{\left( -2q^2P \right) ^{5/4}\left( 3-2\sqrt{-2q^2P} \right)}{\left( 1+\sqrt{-2q^2P} \right) ^{7/2}}.
\end{split}
\end{equation}

By solving Einstein's field equation and the nonlinear Maxwell equation, Ay\'on--Beato and Garc\'ia got an exact four-dimensional stationary spherical charged nonsingular black hole solution, which is called the ABG nonsingular spacetime. The metric can be written in the form~\cite{ABGa98}:
\begin{equation}
		ds^2=-f(r) dt^2
		+f^{-1}(r) dr^2
		+r^2 d\Omega^2,\label{metric}
\end{equation}
where the metric function
\begin{equation}
    f(r)=1-\frac{2 m r^2}{\left(r^2+q^2\right)^{3 / 2}}+\frac{q^2 r^2}{\left(r^2+q^2\right)^2},
\end{equation}
with the non-vanishing components of the electromagnetic tensor:
\begin{equation}
\begin{split}
	F_{rt}=-F_{tr}=\partial_rA_t-\partial_tA_r=q r^4\left(\frac{r^2-5 q^2}{\left(r^2+q^2\right)^4}+\frac{15}{2} \frac{m}{\left(r^2+q^2\right)^{7 / 2}}\right).
\end{split}
\end{equation}
From the definition of the electromagnetic field tensor, we can obtain the nonvanishing components of the vector potential:
    \begin{equation}
       \begin{split}
          A_t&=\int_{\infty}^{r}qr^4\left(\frac{r^2-5q^2}
{(r^2+q^2)^4}+\frac{15}{2}\frac{m}{(r^2+q^2)^{7/2}}\right)dr\\
     &=-\left[\frac{qr^5}{(q^2+r^2)^3}-\frac{3r^5m}{2q(q^2+r^2)^{5/2}}+\frac{3m}{2q}\right],\label{vec potential}
       \end{split}
    \end{equation}
with $m$ representing the ADM mass and $q$ the electric charge of the black hole.

The solution behaves asymptotically as the Reissner-Nordstr\"om spacetime but with a regular center instead of a spacetime singularity~\cite{Bronnikov:2000yz}. The curvature invariants $R$, $R_{\mu\nu}R^{\mu\nu}$, $R_{\mu\nu\alpha\beta}R^{\mu\nu\alpha\beta}$ etc. of the spacetime are regular everywhere~\cite{ABGa98}. The ABG solution can be regarded as a strong field correction to the Reissner-Nordstr\"om solution. This nonlinear electrodynamics satisfies the weak energy condition~\cite{ABGa98}.

The event horizon $\rh$ is the largest positive solution to the equation of the metric function
\begin{equation}
	f(r)=1-\frac{1}{s}\frac{x^2}{(1+x^2)^{3/2}}+\frac{x^2}{(1+x^2)^2}=0,\label{horizons}
\end{equation}
where $x=\rh/|q|$ is a continuous function of the variable $s$. The event horizon of the black hole can be solved directly. It is~\cite{ABGa98}
\begin{equation}
  \begin{split}
     r_{ \pm}  =|q|\left[\left(\frac{1}{4 s}+\frac{\sqrt{h(s)}}{12 s}\pm \frac{\sqrt{6}}{12 s}\sqrt{\frac{9}{2}-12 s^2-\frac{h(s)}{6}
		-\frac{9\left(12 s^2-1\right)}{\sqrt{h(s)}}}\right)^2-1\right]^{\frac{1}{2}},
  \end{split}
\end{equation}
with $h(s)$ and $g(s)$ given by
\begin{align}
 h(s)&=6\left(\frac{3}{2}-4 s^2+s g(s)^{1 / 3}-\frac{4 s\left(11 s^2-3\right)}{g(s)^{1 / 3}}\right), \\
 g(s)&=4\left(9 s+74 s^3+\sqrt{27\left(400 s^6-112 s^4+47 s^2-4\right)}\right).
\end{align}
The number of horizons of the spacetime is related to the parameter $s$. For the extremal ABG black hole, we have:
\begin{equation}
    \begin{split}
        f(\rh)=0, \qquad f'(\rh)=0.
    \end{split}
\end{equation}
Then, we have
\begin{equation}
\begin{split}
	s&= s_0=\frac{t_0(t_0^2-3)}{2t_0^2-4}\approx0.317090,\\
 x&=x_0=\sqrt{t_0^2-1}\approx1.584786,
 \end{split}
\end{equation}
where we have defined a constant
\begin{equation}
\begin{split}
	t_0&=\sqrt{\frac{1}{3}\left( 4+\sqrt[3]{\frac{1}{2}\left( 83-3\sqrt{321} \right)}+\sqrt[3]{\frac{1}{2}\left( 83+3\sqrt{321} \right)} \right)}\approx1.873912
\end{split}
\end{equation}
to simplify the results. The event horizon of the extremal black hole is $r_{\text{ex}}=x_0 |q|$. For $s> s_0$, the spacetime has no horizon; when $s< s_0$, the spacetime has two horizons. For convenience, we assume that the charge of the black hole $q$ is positive.

A black hole is not only a strong gravity system, but also a thermodynamic system, and it has temperature and entropy. The Hawking temperature for the ABG black hole is
\begin{equation}
\begin{split}
 T_{\text{H}}&=\frac{f'(\rh)}{4\pi}=\frac{\rh}{8\pi}\left[\frac{2m}{(\rh^2+q^2)^{3/2}}-\frac{q^2}{(\rh^2+q^2)^2}\right]\frac{\rh^2-2q^2}{\rh^2+q^2}-\frac{q^2\rh^3}{8\pi(\rh^2+q^2)^3}.
	\end{split}
\end{equation}
The Bekenstein-Hawking entropy of the black hole is
\begin{equation}
	S_{\text{BH}}=\pi \rh^2,
\end{equation}
and the electric potential of the event horizon is
\begin{equation}
\begin{split}
	\phi_{\text{h}}(s)=\frac{q\rh^5}{(q^2+\rh^2)^3}-\frac{3\rh^5m}{2q(q^2+\rh^2)^{5/2}}+\frac{3m}{2q}=\frac{x^5-\frac{3}{4 s}x^5\sqrt{1+x^2}}{\left(1+x^2\right)^3}+\frac{3}{4 s}.
\end{split}
\end{equation}

\section{Overcharging the nonsingular black hole with a test particle}\label{particle situation}

As shown in the previous section, the ABG black hole is a nonsingular black hole and without spacetime singularity. Li and Bambi proposed that destroying a nonsingular black hole does not violate the weak cosmic censorship conjecture and may allow us to observe its interior~\cite{Li:2013sea}. Here, we explore the possibility of destroying its event horizon using a charged test particle.

From the divergence of the energy-momentum tensor of the nonlinear electrodynamic field, Liu and Gao obtained a general formula for the general Lorentz-like force and the conserved charge for particles moving in a spacetime with nonlinear electrodynamic field~\cite{LiGa20}.
For a particle with mass $\delta m$ and charge $\delta q$ in spacetime with a nonlinear electrodynamic field,
the conserved quantity associated with a Killing vector field $\xi^{\mu}$ is~\cite{LiGa20}
\begin{equation}
	P=\delta m\xi^{\mu}U_{\mu}+\delta q\xi^{\mu}A_{\mu}.\label{conserved quantity}
\end{equation}
The quantity $P$ is conserved along the trajectory of the motion. Since the test particle does not have magnetic charge, the conserved charge is the same as in linear electrodynamics.

After obtaining the conserved quantities of a charged test particle in a spacetime with a nonlinear electrodynamic field, we check whether the ABG nonsingular black hole can be destroyed by a charged test particle.
We shoot a charged test particle with mass $\delta m$ and charge $\delta q$ into the black hole from rest at infinity. The particle will move along the radial direction, since the angular momentum of the particle is zero. The spacetime is stationary and the energy of the particle is conserved. From the metric of the ABG spacetime, the conserved energy of the electrically charged particle along the worldline is
\begin{equation}
\begin{split}
    \delta E&=-(\delta m~g_{tt}·\dot{t}+\delta q~A_t)\\
    &=\delta m\left(1-\frac{2 m r^2}{\left(r^2+q^2\right)^{3 / 2}}+\frac{q^2r^2}{(r^2+q^2)^2}\right)\frac{dt}{d\tau}+\delta q\left[\frac{r^5(q^2-\frac{3}{2}m\sqrt{r^2+q^2})}{q(q^2+r^2)^3}+\frac{3m}{2q}\right].
\end{split}
\end{equation}

To overcharge the ABG black hole, the particle should be captured by the black hole and cross the event horizon. The worldline of the particle should be timelike and future directed.

Since the four-velocity of the particle is a timelike unit vector, we have
\begin{equation}
g_{\mu\nu}U^{\mu}U^{\nu}=\dot{t}^2g_{tt}+\dot{r}^2g_{rr}=-1.\label{normalization of velocity}
\end{equation}
From the metric of the ABG black hole, we have
\begin{equation}
\begin{split}
	-\dot{t}^2\left(1-\frac{2 m r^2}{\left(r^2+q^2\right)^{3 / 2}}+\frac{q^2r^2}{(r^2+q^2)^2}\right)+\dot{r}^2\left(1-\frac{2 m r^2}{\left(r^2+q^2\right)^{3 / 2}}+\frac{q^2r^2}{(r^2+q^2)^2}\right)^{-1}=-1.
\end{split}
\end{equation}
Then the time component of the four-velocity is
\begin{equation}
\begin{split}
	\dot{t}=\pm\left(1-\frac{2 m r^2}{\left(r^2+q^2\right)^{3 / 2}}+\frac{q^2r^2}{(r^2+q^2)^2}\right)^{-1}\sqrt{\left(1-\frac{2 m r^2}{\left(r^2+q^2\right)^{3 / 2}}+\frac{q^2r^2}{(r^2+q^2)^2}\right)+\dot{r}^2},\label{t's derivative}
\end{split}
\end{equation}
where the sign choice in the solution is dictated by the requirement that the motion of the particle should be future directed, which is $dt/d\tau>0$. Therefore, the energy of the particle is
\begin{small}
\begin{equation}
\begin{split}
	\delta E=\delta m\sqrt{\left(1-\frac{2 m r^2}{\left(r^2+q^2\right)^{3 / 2}}+\frac{q^2r^2}{(r^2+q^2)^2}\right)+\dot{r}^2} +\delta q\left[\frac{r^5(q^2-\frac{3}{2}m\sqrt{r^2+q^2})}{q(q^2+r^2)^3}+\frac{3m}{2q}\right].\label{EofP}
\end{split}
\end{equation}
\end{small}

At the place where the test particle crosses the event horizon, i.e. $r=\rh$, the first term in the right-hand side of eq.~\eqref{EofP} should be non-negative; then we have
\begin{equation}
\begin{split}
	\delta E\geq \delta E_{\text{min}} =\left[\frac{\rh^5(q^2-\frac{3}{2}m\sqrt{\rh^2+q^2})}{q(q^2+\rh^2)^3}+\frac{3m}{2q}\right]\delta q=\phi_{\text{h}}\delta q.\label{Lbound}
 \end{split}
\end{equation}
The result indicates that the energy of the particle should be large enough to overcome the Coulomb force to be captured by the black hole; otherwise, the charged test particle just ``misses'' the black hole due to the Coulomb repulsion force.

After the test particle is captured by the black hole, the spacetime is still described by the metric~\eqref{metric} but the mass $m$ and charge $q$ of the black hole change to $m'=m+\delta E$ and $q'=q+\delta q$, respectively.
To ensure that the event horizon disappears after the particle is captured by the black hole, the energy and charge of the particle should satisfy
\begin{equation}
	s'=\frac{q'}{2m'}=\frac{q+\delta q}{2(m+\delta E)}> s_0.\label{destroyC}
\end{equation}
The equation can be simplified as
\begin{equation}
    \delta E<\delta E_{\text{max}}=\frac{\delta q}{2s_0}+m\left( \frac{s}{s_0}-1\right).\label{Uppbound}
\end{equation}
The result shows that to destroy the event horizon of the ABG black hole, the charge of the test particle should be large enough.

Hence, only the energy $\delta E$ and charge $\delta q$ of the particle satisfy eqs.~\eqref{Lbound} and~\eqref{Uppbound} simultaneously, can the event horizon of the nonsingular ABG black hole be destroyed by the charged test particle.

For an extremal ABG nonsingular black hole, we have $q=2ms_0$ and $x=x_0$. Then
\begin{align}
    \delta E_{\text{min}}&=\phi_{\text{h}}(s_0)\delta q, \label{LowerBoundE} \\
    \delta E_{\text{max}}&=\frac{\delta q}{2s_0}.\label{UperBoundE}
\end{align}
Evaluating the difference between the upper and lower bound of energy, we have
\begin{equation}
\begin{split}
	\delta E_{\text{max}}-\delta E_{\text{min}}=\left(\frac{1}{2 s_0}-\phi_{\text{h}}(s_0)\right)\delta q \approx0.003975\delta q>0.\label{max-min}
\end{split}
\end{equation}
The result shows that there are charged test particles with energy $\delta E$ and charge $\delta q$ that satisfy the two conditions~\eqref{Lbound} and~\eqref{Uppbound} simultaneously. It demonstrates that the event horizon of an extremal ABG black hole can be destroyed by a charged test particle. The
associated interval for mass to charge ratio is relatively narrow, with an order of magnitude of approximately $10^{-3}$. This is quite different from the usual result that an extremal Reissner-Nordstr\"om black hole cannot be destroyed by test charged particles~\cite{Hubeny:1998ga}.

For a near-extremal ABG nonsingular black hole, we define a positive infinitesimal parameter $\epsilon$ to characterize the deviation from an extremal black hole
\begin{equation}
    \epsilon=s_0-s.
\end{equation}
The metric~\eqref{metric} describes an extremal ABG black hole when $\epsilon=0$ and a near-extremal ABG black hole when $0<\epsilon\ll 1$.

To destroy the event horizon of a near-extremal
ABG nonsingular black hole, the charge of the test particle should be large enough. To ensure that the horizon can be destroyed after capturing the charged test particle, the energy and charge of the particle should satisfy eq.(\ref{Lbound}) and eq.(\ref{Uppbound}) simultaneously. For a near-extremal ABG nonsingular black hole, we have
\begin{equation}
   \frac{1}{2s_0}-\phi_{\text{h}} =0.003975+\mathcal{O}(\epsilon).
\end{equation}
It is evident that there are charged test particles with energy $\delta E$ and charge $\delta q$ satisfying eq.~(\ref{Lbound}) and eq.~(\ref{Uppbound}) simultaneously.
So, as with the case of an extremal black hole, a near-extremal ABG nonsingular black hole can be destroyed by a charged test particle.

The above investigation shows that when the nonlinear electrodynamic field effects are taken into account, both extremal and near-extremal ABG nonsingular black holes can be overcharged by a charged test particle. This is different from the gedanken experiment to destroy the Reissner-Nordstr\"om black hole with test charged particles. An extremal charged ABG black hole can also be overcharged by test charged particles at linear order.

\section{Overcharging the nonsingular black hole with a complex scalar field}\label{scalar situation}

In addition to exploring the potential for destroying the event horizon of the charged ABG nonsingular black hole using a test particle, we also investigate the possibility of destroying the event horizon with a charged scalar field. In this section, we examine the possibility of destroying the event horizon of the nonsingular black hole with a charged scalar field.

\subsection{A massive complex scalar field in the ABG spacetime}

We now consider the scattering of a massive complex scalar field in the ABG spacetime. The action for a massive complex scalar field $\Psi$ with mass $\mu_s$ and charge $Q$ minimally coupled to gravity is
\begin{equation}
    S_{\psi}=\int \mathcal{L}_{\Psi} \sqrt{-g}\,\dd^4x,
\end{equation}
with the Lagrangian density $\mathcal{L}_{\Psi}$
\begin{equation}
    \mathcal{L}_{\Psi}=-\frac{1}{2}\mathcal{D}_\mu\Psi\left(\mathcal{D}^\mu\Psi\right)^{*}-\frac{1}{2}\mu_{\text{s}}\Psi\Psi^{*},\label{actionSC}
\end{equation}
where $\mathcal{D}_\mu$ is the covariant derivative $\mathcal{ D}_\mu=\partial_\mu-iQA_\mu$.
The equation of motion for the complex scalar field can be derived from the action. It is
\begin{equation}
	\left(\nabla_{\mu}-iQA_{\mu}\right)\left(\nabla^{\mu}-iQA^{\mu}\right)\Psi-\mu_s^2\Psi=0.
\end{equation}
Expanding the above equation, we obtain
\begin{equation}
	\frac{1}{\sqrt{-g}}\left(\partial_{\mu}-iQA_{\mu}\right)
 \left[\sqrt{-g}g^{\mu\nu}(\partial_{\nu}-iQA_{\nu})\Psi\right]-\mu_s^2\Psi=0.\label{eq of motion}
\end{equation}
Since the spacetime is static and spherically symmetric, the complex scalar field can be decomposed into the following form
\begin{equation}
	\Psi(t,r,\theta,\phi)=e^{-i\omega t}R_{lm}(r)Y_{lm}(\theta,\phi),\label{field}
\end{equation}
where $Y_{lm}(\theta,\phi)$ is the spherical harmonic function.
Substituting eq.~(\ref{field}) into eq.~(\ref{eq of motion}), then the equation of motion for the complex scalar field can be decomposed into the radial part and the angular part. The radial part of the equation is
\begin{equation}
  \frac{1}{r^2}\frac{d}{dr}\left[r^2f(r)\frac{dR_{lm}}{dr}\right]+\left[\frac{(\omega+QA_t)^2}{f(r)}-\frac{l(l+1)}{r^2}-\mu_s^2\right]R_{lm}=0,
\end{equation}
where $l(l+1)$ is the eigenvalue of the spherical harmonic function and $l$ is a positive integer. The solution to the angular part of the equation is the spherical harmonic function. Given that the angular solution is well established and can be normalized to unity in the calculation of the energy and charge fluxes, our focus is on the radial part of the equation.

To solve the radial equation, we define the tortoise coordinate
\begin{equation}
	\frac{dr}{dr_*}=f(r).
\end{equation}
When $r$ varies from the event horizon $r_{\text{h}}$ to infinity $\infty$, the tortoise coordinate $r_*$ ranges from $-\infty$ to $\infty$, and therefore covers the entire spacetime outside the event horizon. After the transformation,
the radial equation can be simplified as
\begin{equation}
  \frac{d^2R_{lm}}{dr_*^2}+\frac{2f(r)}{r}\frac{dR_{lm}}{dr_*}+\left[(\omega+QA_t)^2 -f(r)\left(\frac{l(l+1)}{r^2}-\mu_s^2\right)\right]R_{lm}=0.\label{radial eq}
\end{equation}
Near the event horizon, we have  $r \approx\rh$ and $f(\rh)=0$.  Equation~(\ref{radial eq}) can be approximated as
\begin{equation}
	\frac{d^2R_{lm}}{dr_*^2}+(\omega-Q\phi_{\text{h}})^2R_{lm}=0.\label{radial eq after approx}
\end{equation}
The above equation represents a harmonic oscillatory system and can be solved directly. The solution to eq.~\eqref{radial eq after approx} is
\begin{equation}
	R_{lm}(r)\approx  \exp\left[\pm i(\omega-Q\phi_{\text{h}})r_*\right].
\end{equation}
The solution with a positive sign on the exponent corresponds to the outgoing wave modes, while a negative sign on the exponent corresponds to the ingoing wave modes. It is appropriate to select the negative sign since the ingoing wave mode is a physically acceptable solution near the horizon. Consequently, the charged complex scalar field near the event horizon has the following simple form
\begin{equation}
	\Psi\approx e^{-i(\omega-Q\phi_{\text{h}})r_*}Y_{lm}(\theta,\phi)e^{i\omega t}.
\end{equation}
Once the complex scalar field has been obtained, it is possible to calculate the energy and charge fluxes crossing the black hole event horizon.

\subsection{Destroying the ABG black hole with a complex scalar field}

Since the ABG black hole is nonrotating, we shoot a complex scalar field with frequency $\omega$ and azimuthal angular momentum $m=0$ into the black hole. The variation of parameters of the black hole can be calculated from the fluxes of the charged scalar field during the scattering process.

The Lagrangian for the test complex scalar field is invariant under $t$-translation and $U(1)$ transformation. Correspondingly, there are two conserved currents for these two continuous symmetries. The energy density current and electric density current are the corresponding conserved currents. A straightforward application of Noether's theorem gives the energy density current~\cite{Srednicki:2007qs}:
\begin{equation}
    \begin{split}
         \mathcal{E}^{\mu}=-\frac{\partial \mathcal{L}}{\partial(\partial_\mu\Psi)}\partial_t\Psi-\frac{\partial \mathcal{L}}{\partial(\partial_\mu\Psi^*)}\partial_t\Psi^*=\frac{1}{2}\mathcal{D}^{\mu}\Psi\partial_{t}\Psi^*+\frac{1}{2}(\mathcal{D}^{\mu}\Psi)^*\partial_{t}\Psi.
    \end{split}
\end{equation}
Similarly, the charge density current of the complex scalar field is~\cite{Torres:2014fga}
\begin{equation}
    j^{\mu}=-\frac{1}{2}iQ\left(\Psi^*\mathcal{D}^{\mu}\Psi-\Psi(\mathcal{D}^{\mu}\Psi)^*\right).\label{ecurrent}
\end{equation}
The energy flux through the event horizon is
\begin{equation}
    \begin{split}
    \frac{d E}{dt}=\int_{\text{H}}\mathcal{E}^r\sqrt{-g}\,\dd\theta\, \dd\phi =\omega(\omega-Q\phi_{\text{h}})\rh^2,\label{energy flux}
    \end{split}
\end{equation}
and the electric flux through the event horizon is
\begin{equation}
    \begin{split}
	\frac{dq}{dt}=-\int_{\text{H}}j^r\sqrt{-g}\,\dd\theta\, \dd\phi=Q(\omega-Q\phi_{\text{h}})\rh^2,\label{charge flux}
    \end{split}
\end{equation}
where we have used the normalization condition for the spherical harmonic functions $Y_{lm}(\theta,\phi)$ in the calculation. From eqs.~\eqref{energy flux} and~\eqref{charge flux}, it is clear that when the frequency of the complex scalar field $\omega<Q\phi_{\text{h}}$, the energy and charge fluxes are negative. This means that energy and charges are extracted from the black hole. This is called superradiance~\cite{Brito:2015oca}.

During the scattering process, the energy and charge absorbed by the black hole are
\begin{align}
	dm&=dE=\omega(\omega-Q\phi_{\text{h}})\rh^2\,dt,\label{dif of m}\\
	dq&=Q(\omega-Q\phi_{\text{h}})\rh^2\,dt.\label{dif of q}
\end{align}
The ratio of the energy $dE$ and charge $dq$ absorbed by the black hole is
\begin{equation}
    \frac{dE}{dq}=\frac{\omega}{Q},\label{dE/dq}
\end{equation}
which is consistent with Benkenstein's argument~\cite{Bekenstein:1973mi}.
During the scattering process, the black hole absorbs the complex scalar field and its parameters change to
\begin{equation}
    \begin{split}
        m\rightarrow m'=m+dm, \qquad q\rightarrow q'=q+dq,
    \end{split}
\end{equation}
and the parameter $s$ changes to $s+\delta s$
\begin{equation}
	\delta s=\frac{q+dq}{2\left(m+dm\right)}-s=-\frac{q}{2m^2}dE+\frac{1}{2m}dq.\label{s'}
\end{equation}
Substituting eqs.~(\ref{dif of m}) and~(\ref{dif of q}) into eq.~(\ref{s'}), we have
\begin{equation}
\begin{split}
     \delta s&=-\frac{q}{2m^2}\omega(\omega-Q\phi_{\text{h}})\rh^2\,dt+\frac{1}{2m}Q(\omega-Q\phi_{\text{h}})r_{h}^2\,dt\\
    &=-\frac{Q^2}{m}s\left(\frac{\omega}{Q}-\frac{1}{2s}\right)\left(\frac{\omega}{Q}-\phi_{\text{h}}\right)\rh^2\,dt.\label{change of s}
\end{split}
\end{equation}
Considering an initial extremal black hole, which is characterized by $s=s_0$. After the scattering of the complex scalar field by the extremal black hole, the parameter $s'$ of the final composite object is
\begin{equation}
	 s'=s_0-\int\frac{Q^2}{m}s_0\left(\frac{\omega}{Q}-\frac{1}{2s_0}\right)\left(\frac{\omega}{Q}-\phi_{\text{h}}\right)\rh^2\,\dd t.
\end{equation}
The electrostatic potential of the extremal black hole is
\begin{equation}
\begin{split}
    \phi_{\text{h}}=\frac{x_0^5-\frac{3}{4 s_0}\sqrt{1+x_0^2}x_0^5}{(1+x_0^2)^3}+\frac{3}{4 s_0}\approx 1.572935 <\frac{1}{2s_0}\approx1.576839.
     \end{split}
\end{equation}
If we shoot the complex scalar field with frequency
\begin{equation}
	\phi_{\text{h}}<\frac{\omega}{Q}<\frac{1}{2s_0},\label{Einterval}
\end{equation}
then we have $s'-s_0>0$. In this case, the event horizon disappears. Hence, the extremal ABG black hole can be overcharged.

As can be observed, the widths of the interval are analogous to the particle situation
\begin{equation}
	\frac{1}{2s_0}-\phi_{\text{h}}\approx0.003975,
\end{equation}
the allowed window for the ratio of energy and charge for the complex scalar field $\omega/Q$ to destroy an extremal ABG black hole is quite small.
Similar to the case of a charged particle, the energy-to-charge ratio interval~\eqref{Einterval} for a complex scalar field to destroy the extremal ABG black hole matches that of the charged particle case~\eqref{max-min}.

Suppose that the electromagnetic field is linear, then the metric~\eqref{metric} reduces to the metric of Reissner-Nordstr\"om spacetime. The parameters of the extremal black hole become
\begin{equation}
	s_0=\frac{q}{2m}=\frac{1}{2}, \qquad x_0=\frac{\rh}{q}=\frac{\rh}{m}=1,
\end{equation}
and the electrostatic potential of the extremal Reissner-Nordstr\"om black hole is
\begin{equation}
    \phi_{\text{h}}=\frac{q}{m}=\frac{1}{2s_0}.
\end{equation}
The result shows that the event horizon of an extremal Reissner-Nordstr\"om black hole cannot be destroyed by a charged complex scalar field. It is consistent with previous research that an extremal Reissner-Nordstr\"om black hole cannot be destroyed by a complex scalar field~\cite{Gwak:2021tcl}. In addition, the results show that because of the nonlinear electrodynamic field correction, the extremal spherical charged nonsingular black hole can be destroyed.


\section{Conclusion and Discussion}\label{4section}

The weak cosmic censorship conjecture is one of the cornerstones of black hole physics.  The destruction of a nonsingular black hole does not contravene the weak cosmic censorship conjecture. Consequently, it is possible to destroy a nonsingular black hole in principle.

In this paper, we studied the possibility of destroying an electrically charged nonlinear electrodynamic nonsingular black hole in Einstein gravity. Due to the nonlinear electrodynamic field, the nonsingular ABG spacetime differs from the Reissner-Nordstr\"om spacetime in the strong field region. We investigated the possibility of destroying extremal and near-extremal ABG black holes with a test particle and a complex scalar field and explored the effects of the nonlinear electrodynamic field on the destruction of the event horizon. We identified the parameter ranges that enable the destruction of the event horizon in extremal and near-extremal ABG black holes. Unlike the Reissner-Nordstr\"om black hole, we found that not only a near-extremal but also an extremal ABG black hole can be overcharged by a test charged particle. For the scattering of the complex scalar field, we found that the test field could destroy an extremal ABG black hole. It is truly inspiring that the parameter region for the energy-to-charge ratio, which leads to the destruction of the event horizon by a charged scalar field in an extremal ABG black hole, coincides with that of a test charged particle.

In our study, we observed that while the energy-to-charge ratio of a test scalar field and a test charged particle required to destroy the event horizon is the same, the parameter space for energy and charge is distinct. Numerous studies have explored the destruction of black hole event horizons using thought experiments involving test particles and scalar fields. However, the results for test particles and scalar fields often differ (See refs.~\cite{Jacobson:2009kt,Gwak:2018akg}). At linear order, test particles can destroy a near-extremal black hole, but not an extremal one, whereas test scalar fields cannot destroy either extremal or near-extremal black holes. The fact that both the scalar field and test particle yield the same threshold conditions may be a special case, possibly specific to ABG black holes or nonlinear electrodynamics-corrected nonsingular black holes.

Since the nonsingular ABG spacetime behaves asymptotically to the Reissner-Nordstr\"om spacetime, the nonsingular ABG black hole can be regarded as a strong field correction to the Reissner-Nordstr\"om black hole. In contrast to the finding that neither a test charged particle nor a charged scalar field can destroy an extremal Reissner-Nordstr\"om black hole~\cite{Hubeny:1998ga,Gwak:2021tcl}, we demonstrate that both a test particle and a charged scalar field can overcharge an extremal nonsingular ABG black hole at linear order, owing to nonlinear electrodynamic field corrections. In our analysis, we ignored the radiative and self-force effects. The range of parameters that would allow for the destruction of the event horizon remains quite narrow, on the order of $10^{-3}$. However, when more realistic effects, such as radiative and self-force effects~\cite{Barausse:2010ka}, are included, it is possible that neither a charged test particle nor a test complex scalar field can destroy the event horizon. Nevertheless, with the inclusion of quantum corrections, the black hole may still be susceptible to being destroyed~\cite{Matsas:2007bj,Hod:2008zza,Yang:2025leh}. There is, however, no consensus on this issue as of yet~\cite{Ong:2020xwv}.

The investigation of destroying the event horizon of a nonsingular black hole in general relativity by taking into account the nonlinear electrodynamic field effects, might give us new insight into the weak cosmic censorship conjecture and the possibility of destroying nonsingular black holes. It might also provide a means to explore a black hole's interior~\cite{Li:2013sea}. An event horizon is a null hypersurface where signals from the interior cannot escape to the outside. If the event horizon of a regular black hole is destroyed, the previously hidden interior of the black hole becomes accessible to external observers. This would allow for direct observation of the signals from the interior of what was once a black hole, including shadows and gravitational waves. While previous studies of non-singular black holes have considered the effects of a nonsingular black hole on gravitational waves and electromagnetic signals~\cite{Riaz:2022rlx}, the destruction of the event horizon suggests a more direct way to explore their interiors. This could provide new observational signatures about quantum gravity, potentially revealing details about the underlying spacetime structure, the nature of the central region, deviations from classical general relativity predictions at strong curvature, probably the fate of matter that once fell into the black hole. Furthermore, Hawking’s proof of the black hole area theorem, which establishes an upper bound on the energy released during the collision of two black holes, is based on the weak cosmic censorship conjecture~\cite{Hawking:1971tu}. As noted by Bambi et al.~\cite{Li:2013sea}, the destruction of a regular black hole would violate Hawking’s area theorem. In this case, the concept of an upper bound on the energy released would not apply to a regular black hole. This could be tested through future gravitational wave and astrophysical observations.
\acknowledgments

The authors thank Yu-Xiao Liu for insightful suggestions and helpful discussions. We extend our sincere gratitude to the anonymous reviewers for their insightful comments that have substantially enhanced the quality of this work. This work was supported by the National Natural Science Foundation of China (Grants No. 12305065, No. 12247178, and No. 12247101), the China Postdoctoral Science Foundation (Grant No. 2023M731468), the Fundamental Research Funds for the Central Universities (Grant No. lzujbky-2024-jdzx06), the Natural Science Foundation of Gansu Province (No. 22JR5RA389), and the `111 Center' under Grant No. B20063.

\providecommand{\href}[2]{#2}\begingroup\raggedright\endgroup
\end{document}